\documentclass[11pt]{article}
\newcommand{\be}{\begin{equation}}
\newcommand{\ee}{\end{equation}}
\newcommand{\beq}{\begin{eqnarray}}
\newcommand{\eeq}{\end{eqnarray}}
\newcommand{\beqs}{\begin{eqnarray*}}
\newcommand{\eeqs}{\end{eqnarray*}}

\begin{document}
\noindent
{\Large \bf Holographic dark energy and the universe expansion}
\vskip 0.1cm
\noindent
{\Large \bf acceleration}
\vskip 0.7cm
\noindent
{\bf J.~P.~Beltr\'an Almeida and J.~G.~Pereira} \\
{\it Instituto de F\'{\i}sica Te\'orica} \\
{\it Universidade Estadual Paulista} \\
{\it Rua Pamplona 145} \\
{\it 01405-900 S\~ao Paulo, Brazil}

\vskip 0.8cm
\begin{abstract}
\noindent
By incorporating the holographic principle in a time-depending $\Lambda$-term cosmology, new physical bounds on the arbitrary parameters of the model can be obtained. Considering then the dark energy as a purely geometric entity, for which no equation of state has to be introduced, it is shown that the resulting range of allowed values for the parameters may explain both the coincidence problem and the universe accelerated expansion, without resorting to any kind of additional structures.
\end{abstract}

\section{Introduction}

The observed acceleration in the universe expansion rate \cite{riess-perlmutter,cmb} is usually attributed to the presence of an exotic kind of energy, called dark energy. A great variety of dark energy models have been proposed \cite{5essence,phantom}, but most of them are not able to explain all features of the universe, or are artificially constructed in the sense that it introduces too many free parameters to be able to fit with the experimental data, like for example the {\it coincidence problem}. On the other hand, there is the cosmological-term model, sometimes considered as the most simple and natural of all models \cite{pady1}. Recently, a further development of this model was proposed in which the dark energy associated to a cosmological term $\Lambda$ obeys the so called holographic principle. This principle, originally formulated by 't Hooft and Susskind \cite{hologpp}, states that the number of degrees of freedom directly related to the entropy scales, not with the volume, but with the area of the surface enclosing the system. For applications in cosmology, the principle has been reformulated in a more appropriate form, which uses the light-cone surfaces as boundaries \cite{holog-cosmo, light cones}. To eliminate redundant degrees of freedom, Cohen {\it el al} \cite{cohen} proposed a relationship between the infrared (IR) and the ultraviolet (UV) cutoff, in such a way that the energy density of the system should obey $\epsilon  \leq {\kappa}^{-2} L^{-2}$, where $\kappa^2={8\pi G}/{3c^4}$ and $L$ the IR cutoff parameter. More recently, based on this idea, Li \cite{li} proposed a model with an area-scaling bound for the dark energy density, and discussed several choices for the IR cutoff parameter. Subsequently, by assuming an interacting dark energy scenario, Pav\'on and Zimdhal \cite{pav-zim} considered the case in which the IR cutoff parameter was given by the inverse Hubble parameter. They found that this choice could solve the cosmic coincidence problem, provided an appropriate equation of state for dark energy was used.

In the present letter, we consider a cosmological term model in which we interpret dark energy, not as a real fluid, but as a purely geometric entity. In this case, no equation of state for the dark energy has to be introduced \cite{he}. This assumption is based on the remarkable fact that Einstein's equation has a {\it sourceless} solution with non-vanishing curvature, namely the de Sitter solution. That it is not related to any matter source can be seen by the fact that it is not asymptotically flat. Since a curved spacetime has an intrinsic energy density, this means that energy can be stored by spacetime itself. This is the kind of energy which is usually called dark energy, and for which no source is necessary in Einstein's equation. In addition, we use the holographic model as an extra physical principle ruling the dynamics of a universe with with an interacting dark energy. The main difference in relation to the previous models is that we assume a time-dependent cosmological-term \cite{decay}. A crucial point is to observe that, on account of the Bianchi identity, the sum of the matter plus dark energy is conserved in the covariant sense only. This means that, for a time-varying cosmological term, dark energy can be transformed into ordinary matter, and {\it vice versa}. New holographic bounds for the cosmological parameters are then found, whose range of allowed values may explain both the coincidence problem \cite{coincidence} and the accelerated regime of the universe \cite{riess-perlmutter,cmb}, without resorting to any kind of additional exotic structures.

\section{Holography and $\Lambda$-decaying models}
  
A time-decaying cosmological term is consistent with general relativity, provided matter and/or
radiation is created to make overall energy conserved \cite{abp1}. To see that, let us consider the ``standard'' FRLW equations for a universe with a time-varying cosmological term and flat spatial sections,
\begin{eqnarray}
&{}&\dot{a}^2 = \kappa^2 c^2 [\epsilon_m + \epsilon_\Lambda]a^2 \label{1} \\
&{}&\ddot{a} = \kappa^2 c^2 [\epsilon_\Lambda-\frac{1}{2}(1+3\omega)\epsilon_m]a \label{2}\\
&{}&\dot{\epsilon_m} + 3H(1+\omega)\epsilon_m = - \dot{\epsilon}_\Lambda, \label{3} 
\end{eqnarray}
where $\epsilon_\Lambda$ is the energy density associated to the cosmological term, $a=a(t)$ is the scale factor, and $H={\dot{a}}/{a}$ is the Hubble parameter, with the ``dot'' denoting a time derivative. The last equation is the total energy conservation. In these equations, the matter content of the universe has already been supposed to satisfy the equation of state
\[
p_m = \omega \, \epsilon_m,
\]
where $p_m$ is the pressure, $\epsilon_m$ is the matter energy density, and $\omega$ is a
parameter that depends on the kind of matter.

Now, as is well known, the system of equations (\ref{1}-\ref{3}) has actually only two
independent equations. However, since the cosmological term is time-dependent, there are
three independent dynamical variables. This means that an extra equation, or an extra principle,
will be necessary to solve it.  A way out from this problem is to use the holographic dark energy model, according to which the dark energy density scales with the area of the surface enclosing the system, 
\be
\epsilon_\Lambda\leq b \, \kappa^{-2} \, L^{-2},
\label{elam}
\ee
where $b$ is a free dimensionless parameter, and $L$ is a characteristic length-parameter of the system. Of course, in order to have a positive dark energy density, the condition $b>0$ must be satisfied.

The next step is to determine the parameter $L$. A natural choice is to identify it with the inverse Hubble radius \cite{pav-zim}: $L = cH^{-1}$. In addition to being the simplest one, it yields a dark energy density which is comparable to the present day value \cite{thomas}. In this case, the condition (\ref{elam}) can be rewritten in the form
\be
\label{boundH}
\epsilon_\Lambda \leq b \, c^{-2} \kappa^{-2} \, H^{2}.
\ee
Using the fact that $b$ is a free parameter, it is possible to saturate the above inequality by imposing restrictions on $b$. In this way an additional equation is obtained which allows the system of equations (\ref{1}-\ref{3}) to be solved in terms of the holographic parameter $b$. The solution for the matter and the dark energy densities are, respectively,
\be
\epsilon_m = \alpha a^{-3(1+\omega)(1-b)}
\ee
and
\be
\epsilon_\Lambda= \frac{b}{1-b}\alpha a^{-3(1+\omega)(1-b)},
\ee 
with $\alpha$ an integration constant. They are easily found to satisfy the relation
\be
\label{m-d}
\epsilon_m = \frac{1-b}{b} \epsilon_\Lambda.
\ee 

Now, if we want to preserve the positivity of $\epsilon_m$, we see from Eq.\ (\ref{m-d}) that $b<1$, a result that agrees with the N-bound argument \cite{N-bound}. In fact, according to this argument, the maximum of entropy in an asymptotically de Sitter spacetime is achieved in the pure de Sitter case. The maximum of the energy density, therefore, will also be achieved in the pure de Sitter spacetime, where the natural length-parameter $L$ is the de Sitter radius $l$. In this case, as follows from Eq.~(\ref{elam}),
\be
\epsilon_\Lambda\leq b \, \kappa^{-2} \, l^{-2}.
\ee
Since $\kappa^{-2} \, l^{-2}$ represents the dark energy density of a pure de Sitter spacetime, which is the case of highest energy energy density, we see that $b<1$. The holographic parameter $b$, therefore, must be in the interval
\be
0 < b < 1.
\ee
For an appropriate value of the parameter $b$ within this interval, the relation (\ref{m-d}) can yield the experimentally observed relation between the densities, known as the coincidence problem.

For the sake of completeness, we give below the time evolution of the Hubble parameter and of the scale factor:
\be
\label{H}
H = \frac{H_0}{1+\frac{3}{2}H_0(1+\omega)(1-b)(t-t_0)}
\ee
and
\be
\label{a}
a = a_0 \Big[1+\frac{3}{2} H_0(1+\omega)(1-b)(t-t_0) \Big]^{{2}/[3(1+\omega)(1-b)]}.
\ee
The matter and dark energy densities will, consequently, scale as
\be
\epsilon_m\sim \epsilon_\Lambda \sim
\Big[1+\frac{3}{2} H_0(1+\omega)(1-b)(t-t_0) \Big]^{-2}.
\ee

\section{Holographic Bounds}

We analyze now the restrictions imposed by the holographic principle on the parameters of the model, in particular on the arbitrary holographic $b$-parameter and on the $\omega$-parameter of the matter equation of state. We emphasize that we are not imposing any {\it a priori} restriction on the $\omega$-parameter, which is free to assume any value, even in the present era. We are not, therefore, supposing that dust matter rules the present dynamics of the universe. Furthermore, we are assuming that the equations are valid at any epoch, and consequently for any arbitrary fluid (or a mixture of several fluids) interacting with the cosmological term. This means that the bounds we are going to obtain are valid at any epoch of the universe evolution. It should be remarked that the dark energy will be interpreted, not as a fluid, but as a purely geometrical entity, for which no equation of state has to be introduced \cite{he}.

\subsection{Bounds from the energy}

By imposing physically reasonable conditions on the matter--plus--dark energy system, new bounds on the parameters can be obtained. The first set of conditions are the so called energy conditions, which in a time-varying $\Lambda$ model must be applied on the energy-momentum tensor
\be
\Theta_{\mu\nu}= T_{\mu\nu} + \Lambda_{\mu\nu}.
\ee
According to the Bianchi identity, it is covariantly-conserved \cite{abp1}:
\be
\label{e-m cons}
\nabla_\mu\Theta^{\mu}{}_{\nu} = \nabla_\mu [T^{\mu}{}_{\nu} +
\Lambda^{\mu}{}_{ \nu}] = 0.
\ee
Here,
\be
T_{\mu\nu}= \epsilon_m u_\mu u_\nu+p_m(g_{\mu \nu}+u_\mu u_\nu )
\ee
is the matter energy-momentum tensor, and
\be
\Lambda_{\mu\nu}=-\epsilon_\Lambda g_{\mu \nu}
\ee
is the energy-momentum tensor associated to the cosmological term.
 
Using the matter equation of state $p_m = \omega \epsilon_m$, as well as Eq.\ (\ref{m-d}), we can write
\be
\label{e-m}
\Theta_{\mu\nu} = \epsilon_m (1+\omega) u_\mu u_\nu +
\epsilon_m \Big(\omega -\frac{b}{1-b} \Big) g_{\mu \nu}.
\ee
As is well known, the energy conditions are the following \cite{wald}: \\ 
(i) {\it Null energy condition}: for all null vector $n^\nu$,
\be
\Theta_{\mu\nu}n^\mu n^\nu \geq 0,
\ee
which means that light rays are focussed by matter.\\
(ii) {\it Weak energy condition}: for all time-like vector $v^\nu$,   
\be
\Theta_{\mu\nu}v^\mu v^\nu \geq 0.
\ee
(iii) {\it Causal energy condition}: for all time-like vector $v^\nu$,   
\be
\Theta^\mu{}_\nu v_\mu \Theta^\nu{}_\alpha v^\alpha \leq 0,
\ee
which, roughly speaking, means that the energy cannot flow faster than light. 
  
Applied on the energy-momentum tensor (\ref{e-m}), we obtain from the null energy condition that
\be
\omega\geq -1,
\label{ec1}
\ee
which is the usual result of the standard cosmology. On the other hand, from the weak and the causal conditions, we get respectively
\be
\Theta_{\mu\nu}v^\mu v^\nu= \epsilon_m (1+\omega) (u^\mu v_\mu)^2+
\epsilon_m  \Big(\omega -\frac{b}{1-b} \Big) v^2 \geq 0
\ee
and
\be
\Theta^\mu{}_\nu v_\mu \Theta^\nu{}_\alpha v^\alpha =
\epsilon_m^2 \bigg[(1+\omega)^2u^2 (u^\mu v_\mu)^2+
2 (1+\omega) \Big(\omega - \frac{b}{1-b} \Big) (u^\mu v_\mu)^2+
\Big(\omega -\frac{b}{1-b}\Big)^2 v^2 \bigg] \leq 0.
\ee
Given that these relations are valid for any time-like $v^\mu$, both of them yield 
\be
\omega \leq \frac{b}{1-b}.
\label{ec2} 
\ee
This is a new feature of the interacting (that is, dynamic) holographic dark energy model. In fact, in the context of general relativity without holographic dark energy term, the positive energy conditions requires that $\omega < 1$ in order to preserve causality. Therefore, causality is modified by the holographic principle: a new ``causal"  bound is obtained, which is ruled by the holographic $b$-parameter. That a variable cosmological term changes the causal properties of spacetime can be understood from the fact that both particle and future horizons for any observer (which define the boundary of the regions causally connected to the observer) must change as the cosmological constant decays, giving rise to a concomitant matter creation. A related discussion can be found in Ref.~\cite{abp2}.

\subsection{Bounds from the entropy}

There are other bounds and relations between the parameters $b$ and $\omega$ that can obtained from the holographic principle. As an example, we present in this section a calculation along the lines of the holographic principle adapted to homogeneous cosmological solutions \cite{holog-cosmo}. For the particle horizon
\be
R_H(t)=\int_0^t\frac{dt}{a(t)},
\ee 
the holographic principle states that the total entropy $\sigma=\sigma_m + \sigma_\Lambda$ inside the horizon does not exceed the area of the horizon:
\be
\sigma R_H^3< \left(aR_H \right)^2.
\ee
Using the expansion parameter $a$ from equation (\ref{a}), the  principle implies that
\be
\frac{\sigma}{a_0^3}H_0\beta(\beta -1)\left(1+H_0\beta t \right)^{\frac{3(\beta -1)}{\beta} -2}<1,
\ee
where $\beta=3(1+\omega)(1-b)/2$. Now, supposing an entropy density of the form \cite{mbs}
\[
\sigma \sim \epsilon^{{\gamma}/{(1+\omega)}} \sim a^{-3\gamma(1-b)},
\]
with either $\epsilon = \epsilon_m$ or $\epsilon=\epsilon_\Lambda$, and $\gamma>0$ another unknown parameter, we obtain 
\be
\frac{k_0}{a_0^3}H_0\beta(\beta -1) a_0^{-3\gamma(1-b)} (1+H_0\beta t)^{ -2 + \frac{3(\beta -1)}{\beta} - \frac{3\gamma(1-b)}{\beta}}<1,
\ee
where $k_0$ is a known constant written in terms of the other parameters of the model. In the limit of time going to infinity (in order to preserve the bound at this time), we obtain:
\be
\omega< \frac{b}{1-b} +\frac{1}{1-b}+2\gamma.
\ee
Since $\gamma>0$, this $\omega$ bound is less restrictive than the one imposed by the energy conditions, and for this reason it will not be considered.
 
\subsection{Accelerated regime}

By taking the second time-derivative of the scale factor (\ref{a}), and supposing it to be
positive, we obtain
\be
\label{accel}
\omega< \frac{b}{1-b}-\frac{1}{3(1-b)}.
\ee
This is the condition for an accelerated universe expansion. An interesting point is to observe that, in the present case, the parameter $\omega$ can be different from zero. This means that an accelerated regime in the present era can be obtained even for non-dust matter.\footnote{This an important difference in relation to some previous works \cite{li,pav-zim} in which the condition $\omega=0$ has been assumed.} Since $b<1$, the second term on the right-hand side condition (\ref{accel}) is positive, and consequently it will always satisfy the physical bounds obtained from the energy and entropy constraints. The crucial point is to note that the accelerated expansion is not directly produced by the dark energy itself, but by its dynamical and holographic properties. In fact, the condition for an accelerated expansion rate depends on the holographic parameter $b$. We see in this way that, for the parameters within the limits imposed by the energy and entropy bounds, it is possible to obtain an accelerated expansion regime for the universe, as suggested by recent observations. It should be remarked finally that, for the specific case of $b<1/3$, the parameter $\omega$ becomes negative. However, such a situation, which represents an exotic matter with negative pressure, is not necessary to produce an accelerated universe expansion. In fact, for $1/3<b<1$, although $\omega$ is positive, the universe expansion will be accelerated.

\section{Conclusions}

The main result of this letter is that a dynamical dark energy interacting with matter can ``answer" some of the key questions of modern cosmology, provided the elegant assumption of holographic dark energy is adopted. In fact, as we can see from equations (\ref{m-d}) and (\ref{accel}), for some allowed values of the parameters, it is possible to achieve the ``coincidence" in the order of magnitude for the matter and dark energy densities, and at the same time obtain an accelerated regime for the universe expansion. No additional structures, aside from the holographic decaying cosmological term, turns out to be necessary. Furthermore, since the dark energy is interpreted, not as a fluid, but as a purely geometrical entity, no equation of state for it has to be introduced. In such a model, dark energy and  matter can be transformed into each other. In other words, matter can be transformed in geometry, and vice-versa. In particular, it is possible to conceive an empty universe in which all energy is dark: it is given by an empty cone-spacetime, singular at the vertex, transitive under {\it proper} conformal transformations, which may eventually be related to a initial condition for a big bang universe \cite{abp2}.

The important new feature of the holographic, purely geometrical dynamic dark energy model is that it changes the usual causal conditions imposed on the matter equation of state. As a direct consequence, the acceleration in the universe expansion is not directly produced by the dark energy, but by its dynamical and holographic characters. It should be remarked finally that the ideas presented here do not completely solve the coincidence and the universe accelerated expansion problems. They only show the consistency, as well as its potential ability to describe the dynamics of the universe without resorting to any additional structure. Of course, there is still much to be done towards a consistent solution that meets all observational data.

\section*{Acknowledgments}
The authors would like to thank R. Aldrovandi for useful discussions. They would like to thank also CAPES, CNPq and FAPESP for partial financial support.

\end{document}